\documentclass[twocolumn,twoside,slac_two]{revtex4-1}
\usepackage{graphicx}
\usepackage{amsmath}
\usepackage{fancyhdr}
\begin{document}

\title{Constraining Dark Matter Signal from a Combined Analysis of Milky Way Satellites with the Fermi-LAT}


\author{M. Llena Garde}
\email{maja.garde@fysik.su.se}
\affiliation{Department of Physics, Stockholm University, AlbaNova, SE-106 91 Stockholm, Sweden, The Oskar Klein Centre for Cosmoparticle Physics, AlbaNova, SE-106 91 Stockholm, Sweden}
\author{J. Conrad}
\email{conrad@fysik.su.se}
\affiliation{Department of Physics, Stockholm University, AlbaNova, SE-106 91 Stockholm, Sweden, The Oskar Klein Centre for Cosmoparticle Physics, AlbaNova, SE-106 91 Stockholm, Sweden, Royal Swedish Academy of Sciences Research Fellow, funded by a grant from the K. A. Wallenberg Foundation}
\author{J. Cohen-Tanugi}
\email{johann.cohen-tanugi@lupm.in2p3.fr}
\affiliation{Laboratoire Univers et Particules de Montpellier, Universit\'e Montpellier 2, CNRS/IN2P3, Montpellier, France}
\author{On behalf of the Fermi-LAT collaboration, M. Kaplinghat and G. Martinez}
\noaffiliation

\begin{abstract}
Dwarf spheroidal galaxies have a large mass to light ratio and low astrophysical
background, and are therefore considered one of the most promising targets for
dark matter searches in the gamma-ray band. By applying a joint
likelihood analysis, the power of resultant limits in case of no detection can be
enhanced and robust constraints on the dark matter parameter space can be
obtained. We present results from a combined analysis of 10 dwarf spheroidal
galaxies using Fermi-LAT data. Different annihilation channels have been
analyzed and uncertainties from astrophysical properties have been taken into
account.
\end{abstract}

\maketitle

\thispagestyle{fancy}


\section{Introduction}
The Fermi Gamma-ray Space Telescope was launched on June 11, 2008. Its main instrument, the Large Area Telescope (Fermi-LAT), observes the entire sky every $\sim$3 hours (2 orbits) with a field of view covering $\sim$2.4 sr and a sensitive energy range extending from 20MeV to $>$300GeV \cite{Atwood:2009ez}. These properties make the Fermi-LAT an excellent instrument for dark matter (DM) searches. 

One of the leading DM candidates is a weakly interacting massive particle (WIMP). The gamma-ray flux from self-annihilating WIMPs can be expressed as $\phi_{WIMP} (E, \psi) = J(\psi)\times\Phi^{PP}(E)$ (see {\em e.g.}, \cite{Bergstrom:1997fj}, see also \cite{Bergstrom:2000pn} for a review), where $\Phi^{PP}(E)$ is the "particle physics factor" described by
\begin{equation}
\Phi^{PP}(E)=\frac{<\sigma v>}{8\pi m_{WIMP}^{2}}\times N_W(E)
\end{equation}
and $J(\psi)$ is the "astrophysical factor", or J-factor, described by
\begin{equation}
J(\psi)=\int_{l.o.s.}dl(\psi)\rho^{2}(l(\psi)).
\end{equation}
Here $\left\langle \sigma  v \right\rangle$ is the (velocity-averaged) WIMP annihilation cross section times relative velocity, $m_{WIMP}$ is the WIMP mass, $N_W(E)$ is the gamma-ray energy distribution per annihilation, and $\rho(r)$ is the dark matter density distribution. 

%The particle physics factor has two spectral features: the continuum feature and the line feature.

Dwarf spheroidal galaxies (dSphs) are considered to be DM dominated systems since they have a very high mass to light ratio. They are near-by ($\sim 100 kpc)$ and they have low background since most dSphs are expected to be free from other astrophysical gamma-ray sources and they have a small gas content. This makes them interesting targets for gamma-ray DM searches. However, the expected flux from DM annihilation or decay for dSphs are expected to be very low. The Fermi-LAT collaboration has recently presented results from a DM search in a number of dSphs \cite{Abdo:2010ex}. In this work we perform a joint likelihood analysis, taking advantage of the fact that the DM spectra are the same in all targets.

We present results considering ten dSphs taking into account the uncertainties in the astrophysical factors. The results presented in these proceedings are updated with respect to the preliminary results presented at the conference. This work was initially presented in \cite{Garde:2011wr}, and final results are presented in \cite{collaboration:2011wa}.

\section{Analysis}

\begin{table}[ht]%[H] add [H] placement to break table across pages
\caption{Position, distance, and J-factor (under assumption of a Navarro-Frenk-White profile) of each dSph. The 4th column shows the mode of the posterior distribution of $\log_{10}J$, and the 5th column indicates its 68\% C.L. error. See the text for further details. The J-factors correspond to the pair annihilation flux coming from a cone of solid angle $\Delta\Omega = 2.4\cdot10^{-4}$ sr. The final column indicates the reference for the kinematic dataset used. \label{dwarfs}} 
{\small \hfill{}
\begin{ruledtabular}
\begin{tabular}{ccccccc}
%Lines of table here ending with \\
Name & l &  b & d & $\overline{\log_{10}({J})}$ & $\sigma$ & ref.\\  %(\times10^{19}\frac{GeV2}{cm5})
     & deg.& deg.& kpc&\multicolumn{2}{c}{$\log_{10}[{\rm GeV}^2 {\rm cm}^{-5}]$} \\\hline
Bootes I & $358.08$ & $\phantom{-}69.62$ &60 &$17.7$ &$0.34$ &\cite{Koposov:2011zi} \\ %$0.16^{+0.35}_{-0.13}$ \\ %
Carina & $260.11$ & $-22.22$ & 101&$18.0$ &$0.13$ &\cite{Walker:2009zp} \\ %$0.06^{+0.20}_{-0.10}$ \\
Coma Berenices & $241.9$ &$\phantom{-}83.6$ &44& $19.0$ & $0.37$ &\cite{Simon:2007dq}\\ %$0.16^{+0.22}_{-0.08}$ \\
Draco & $86.37$ & $\phantom{-}34.72$ &80 &$18.8$ & $0.13$ &\cite{Walker:2009zp}\\ %$1.20^{+0.31}_{-0.25}$ \\
Fornax & $237.1$ & $-65.7$ & 138&$17.7$ & $0.23$ &\cite{Walker:2009zp}\\ %$0.06^{+0.03}_{-0.03}$ \\
Sculptor & $287.15$ & $-83.16$ & 80&$18.4$ &$0.13$ &\cite{Walker:2009zp} \\ %$0.24^{+0.06}_{-1.49}$ \\
Segue 1 & $220.48$ & $\phantom{-}50.42$ & 23&$19.6$ & $0.53$ &\cite{Simon:2010ek} \\ %$2.00^{+5.95}_{-0.60}$ \\
Sextans & $243.4$ & $\phantom{-}42.2$ &86 &$17.8$ & $0.23$ &\cite{Walker:2009zp}\\ %$0.06^{+0.03}_{-0.02}$ \\
Ursa Major II & $152.46$ & $\phantom{-}37.44$ & 32&$19.6$ &$0.40$ &\cite{Simon:2007dq}\\ %$0.58^{+0.91}_{-0.35}$ \\
Ursa Minor & $104.95$ & $\phantom{-}44.80$ & 66&$18.5$ &$0.18$ &\cite{Walker:2009zp} \\ %$0.64^{+0.25}_{-0.18}$ \\
\end{tabular}
\end{ruledtabular}
}
\hfill{}
\end{table}

We have observed ten dSphs, listed in Table \ref{dwarfs}, which is the same set of dwarfs for which annihilation cross section limits were presented in \cite{Abdo:2010ex} with the addition of Carina and Segue 1, using 24 months of data. We have used the diffuse event class which only contains the events with the highest gamma-like confidence, and we have chosen events ranging from 200~MeV to 100~GeV. We used the Fermi-LAT instrument responce function P6$\_$V3$\_$DIFFUSE. Our region of interest (ROI) is a region of 10 degrees radius centered on dSph location. Standard cuts removing Earth albedo photons have been made. The dSphs are modeled as DM point sources using the DMFit package \cite{Jeltema:2008hf} where we consider 100\% annihilation into the \textit{b\={b}}, the $\tau^+ \tau^-$, the $W^+ W^-$, and the $\mu^+ \mu^-$ annihilation channel. The background is modeled according to Fermi-LAT recommendations \cite{modelPage}, and sources within 15 degrees are modeled according to the first year point source catalog \cite{Collaboration:2010ru}. We perform a binned analysis to use both energy and spatial information. The data selection and analysis are performed using the Fermi-LAT analysis package, ScienceTools \cite{STpage}, and the upper limits are obtained using profile likelihood as implemented in the MINUIT processor MINOS \cite{minuit}.

One large uncertainty in indirect DM detection methods arises from the uncertainties in the astrophysical factors, the J-factors. We have included these uncertainties by including the distribution of the J-factors in the likelihood fit, treating them as nuissance parameters. This is the first time J-factor uncertainties are included in this way. In our fits, the parameter of interest is the WIMP annihillation cross-section, $\langle\sigma v\rangle$, and the nuissance parameters are the J-factors, the normalizations of the Diffuse Backgrounds and the normalizations of sources within 5 degrees of the dSphs. 

With this addition, the joint likelihood considered in our analysis becomes:
\begin{align}\label{eq:L}
L(D| \mathbf{p_W},&\{\mathbf{p}\}_i)=\prod_{i} L^{\rm LAT}_{i}(D|\mathbf{p_W}, \mathbf{p}_i)\nonumber\\
 & \times \frac{1}{\ln(10)\,J_i \sqrt{2\pi}\sigma_{i}} e^{-\left(\log_{10}(J_i)-\overline{\log_{10}({J}_i)}\right)^2/2\sigma_{i}^2}\; ,
\end{align}
\noindent where $L^{\rm LAT}_i$ denotes the binned Poisson likelihood that is commonly used in a standard single ROI analysis of the LAT data, $i$ indexes the ROIs, $D$ represents the binned gamma-ray data, $\mathbf{p_W}$ represents the set of ROI-independent DM parameters ($\langle\sigma v\rangle$ and $m_W$), $\{\mathbf{p}\}_i$ are the ROI-dependent model parameters. In this analysis, $\{\mathbf{p}\}_i$ includes the normalizations of the nearby point and diffuse sources and the J-factor, $J_i$. $\overline{\log_{10}(J_i)}$ and $\sigma_i$ are the mean and standard deviation of the distribution of $\log_{10}{(J_i)}$, approximated to be gaussian. The values for the J-factors have been updated since the preliminary work was presented at the conference, and the updated values are found in Table \ref{dwarfs}. A detailed discussion about how the J-factors are obtained are found in \cite{collaboration:2011wa}.

We have performed tests on the coverage of the joint likelihood method and also tests on how the combined limits scale with the number of added targets. In our fits, we constrain the  $\langle\sigma v\rangle$ parameter to be positive. Our tests show that this will make the combined limit scale as better than $1/\sqrt{N}$, where N is the number of added targets. The hardness of the DM spectra will also influence, resulting in a larger improvement for harder spectra. The coverage of our method is good but with slight overcoverage for small signals, i.e. conservative limits.

\section{Results}
\begin{figure}
\center
\includegraphics[width=.5\textwidth]{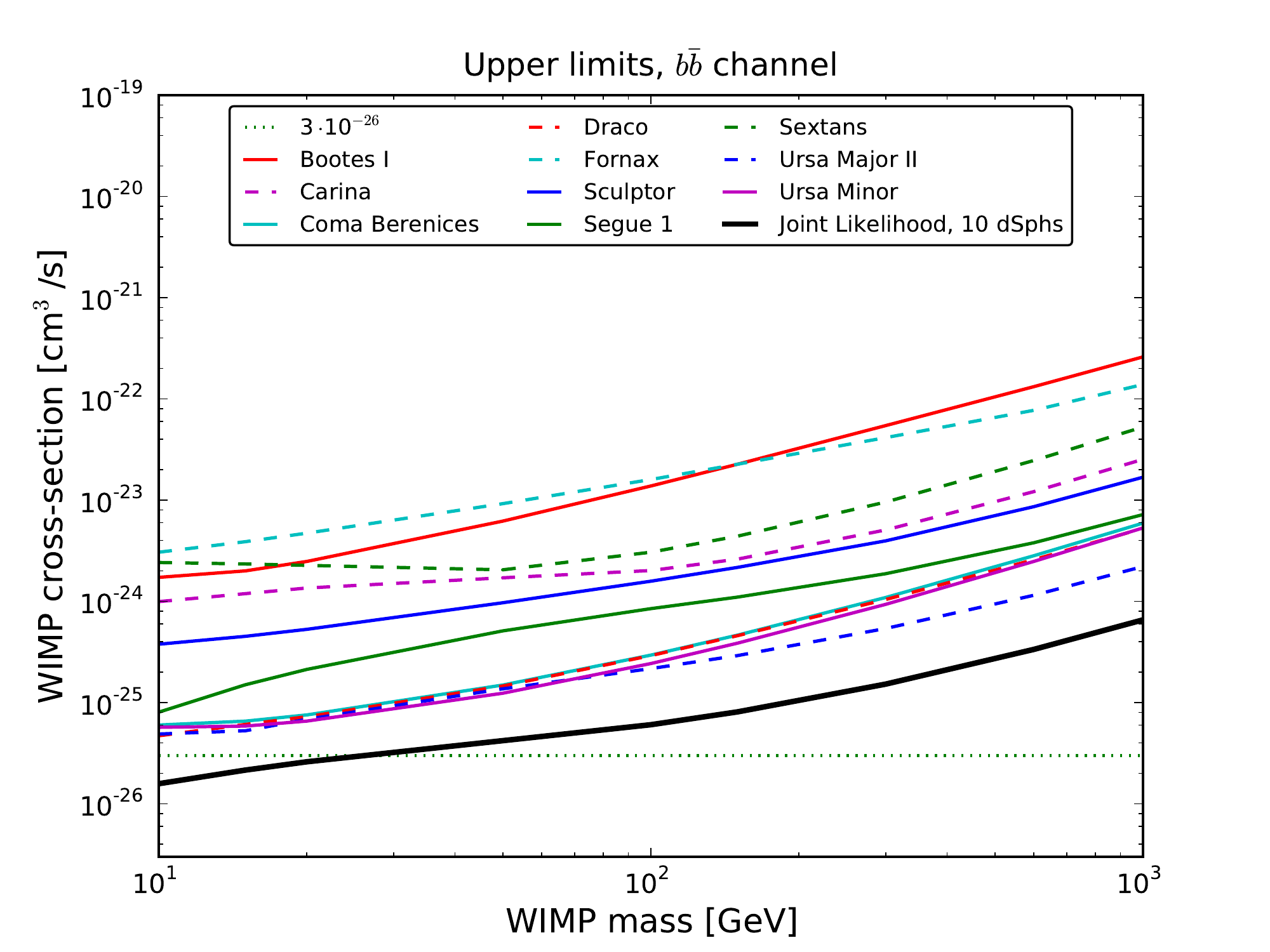}
\caption{95\% Upper limits on WIMP annihilation cross section for annihilation into 100\% \textit{b\={b}}. The expected thermal WIMP cross-section is plotted as a reference. \label{fig1}}
\end{figure}
\begin{figure}
\center
\includegraphics[width=.5\textwidth]{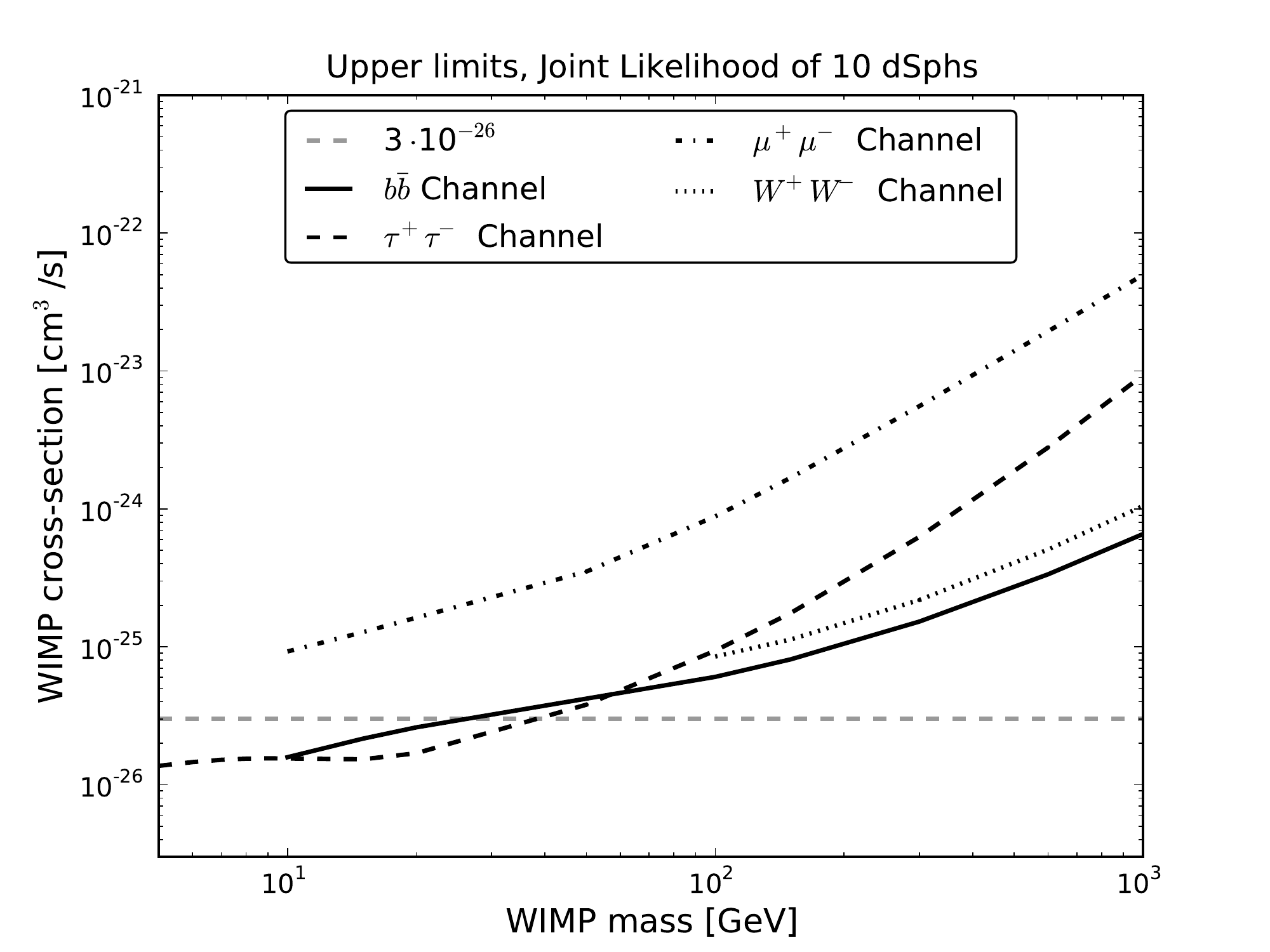}
\caption{95\% Upper limits on WIMP annihilation cross section for annihilation into 100\% \textit{b\={b}}, 100\% $\tau^+ \tau^-$, 100\% $W^+ W^-$, and 100\% $\mu^+ \mu^-$. The expected thermal WIMP cross-section is plotted as a reference. \label{fig2}}
\end{figure}
In Fig. \ref{fig1} we present constraints on the WIMP annihilation cross-section into 100\% \textit{b\={b}} for ten dSphs, both individual and combined limits. In Fig. \ref{fig2} we present the combined limits on WIMP annihilation cross section for annihilation into 100\% \textit{b\={b}}, 100\% $\tau^+ \tau^-$, 100\% $W^+ W^-$, and 100\% $\mu^+ \mu^-$. J-factor uncertainties have been included. The joint likelihood method allows us to rule out WIMP annihilation with cross sections predicted by the most generic cosmological calculation up to a mass of $\sim$30 GeV for the \textit{b\={b}} and the $\tau^+ \tau^-$ annihilation channel.

Final results are presented in \cite{collaboration:2011wa}.

\bigskip % extra skip inserted
\begin{acknowledgments}
The \textit{Fermi} LAT Collaboration acknowledges generous ongoing support
from a number of agencies and institutes that have supported both the
development and the operation of the LAT as well as scientific data analysis.
These include the National Aeronautics and Space Administration and the
Department of Energy in the United States, the Commissariat \`a l'Energie Atomique
and the Centre National de la Recherche Scientifique / Institut National de Physique
Nucl\'eaire et de Physique des Particules in France, the Agenzia Spaziale Italiana
and the Istituto Nazionale di Fisica Nucleare in Italy, the Ministry of Education,
Culture, Sports, Science and Technology (MEXT), High Energy Accelerator Research
Organization (KEK) and Japan Aerospace Exploration Agency (JAXA) in Japan, and
the K.~A.~Wallenberg Foundation, the Swedish Research Council and the
Swedish National Space Board in Sweden.

Additional support for science analysis during the operations phase is gratefully
acknowledged from the Istituto Nazionale di Astrofisica in Italy and the Centre National d'\'Etudes Spatiales in France.
\end{acknowledgments}

\bigskip % extra skip inserted


\begin{thebibliography}{99} % Use for 10-99 references
\bibitem{Atwood:2009ez} W. B. Atwood et al, \emph{The Large Area Telescope on the Fermi Gamma-ray Space Telescope Mission}, \emph{Astrophys. J.} \textbf{697} 1072 (2009).
\bibitem{Bergstrom:1997fj} L. Bergstrom, P. Ullio and J. H. Buckley, \emph{Observability of gamma-rays from dark matter neutralino annihilations in the Milky Way halo}, \emph{Astropart.Phys}, \textbf{9} 137-162 (1998)
\bibitem{Bergstrom:2000pn} L. Bergstrom, \emph{Nonbaryonic dark matter: Observational evidence and detection methods}, \emph{Rept.Prog.Phys.}, \textbf{63} 793 (2000).
\bibitem{Abdo:2010ex} A. A. Abdo et al, \emph{Observations of Milky Way Dwarf Spheroidal galaxies with the Fermi-LAT detector and constraints on Dark Matter models}, \emph{Astrophys. J.} \textbf{712} 147 (2010).
\bibitem{Garde:2011wr} M. Llena Garde for the Fermi-LAT Collaboration, \emph{Constraining dark matter signal from a combined analysis of Milky Way satellites using the Fermi-LAT} Proceedings from IDM 2010 (2011) [{\tt astro-ph.HE 1102.5701}]
\bibitem{collaboration:2011wa} M.Ackermann et al, \emph{Constraining dark matter models from a combined analysis of Milky Way satellites with the Fermi-LAT}, \emph{accepted for PRL} (2011) [{\tt astro-ph.HE 1108.3546}].
\bibitem{Walker:2009zp} Matthew G. Walker et al, \emph{A Universal Mass Profile for Dwarf Spheroidal Galaxies}, \emph{Astrophys. J.}, \textbf{704} 1274-1287 (2009).
\bibitem{Simon:2010ek} Jousha D. Simon et al, \emph{A Complete Spectroscopic Survey of the Milky Way Satellite Segue 1: The Darkest Galaxy}, \emph{Astrophys. J.}, \textbf{733} 46 (2011).
\bibitem{Simon:2007dq} Marla Geha and Joshua D. Simon, \emph{The Kinematics of the Ultra-Faint Milky Way Satellites: Solving the Missing Satellite Problem}, \emph{Astrophys. J.}, \textbf{670} 313-331 (2007).
\bibitem{Koposov:2011zi} Sergey E. Koposov et al, \emph{Accurate Stellar Kinematics at Faint Magnitudes: application to the Bootes~I dwarf spheroidal galaxy}, (2011),[{\tt astro-ph.GA 1105.4102}].
\bibitem{Jeltema:2008hf} Tesla E. Jeltema and Stefano Profumo, \emph{Fitting the Gamma-Ray Spectrum from Dark Matter with DMFIT: GLAST and the Galactic Center Region}, \emph{JCAP} \textbf{0811} 003 (2008).
\bibitem{modelPage} \emph{http://fermi.gsfc.nasa.gov/ssc/data/access/lat/
BackgroundModels.html}.
\bibitem{Collaboration:2010ru} A. A. Abdo et al, \emph{Fermi Large Area Telescope First Source Catalog}, \emph{Astrophys. J. Suppl} \textbf{188} 405 (2010).
\bibitem{STpage} \emph{ http://fermi.gsfc.nasa.gov/ssc/data/analysis/software/}.
\bibitem{minuit} F. James, \emph{MINUIT Reference Manual} CERN Program Library Writeup D506.

%\bibitem{Baltz:2008wd} E. A. Baltz et al, \emph{Pre-launch estimates for GLAST sensitivity to Dark Matter annihilation signals}, \emph{JCAP} \textbf{0807} 013 (2008). 
%\bibitem{LATcaveats} \emph{http://fermi.gsfc.nasa.gov/ssc/data/analysis/
%LAT\_caveats.html}.


\end{thebibliography}
%\begin{thebibliography}{9}   % Use for  1-9  references

\end{document}